%% file: main.tex
\documentclass{ssh}

\input{preamble.tex}
\renewcommand{\st}[1]{}
\title{Imaging Atomic-Scale Chemistry from \\ Fused Multi-Modal Electron Microscopy}

\author[1]{Jonathan~Schwartz}
\author[2]{Zichao~Wendy~Di}
\author[3]{Yi~Jiang}
\author[4]{Alyssa~J.~Fielitz}
\author[5,6]{Don-Hyung~Ha}
\author[5]{Sanjaya~D.~Perera}
\author[7,8]{\\Ismail~El~Baggari}
\author[5]{Richard~D.~Robinson}
\author[9]{Jeffrey~A.~Fessler}
\author[10]{Colin~Ophus}
\author[4]{Steve~Rozeveld}
\author[1,11,*]{Robert~Hovden}

\affil[1]{Department of Materials Science and Engineering, University of Michigan, Ann Arbor, MI}
\affil[2]{Mathematics and Computer Science Division, Argonne National Laboratory, Lemont, IL}
\affil[3]{Advanced Photon Source Facility, Argonne National Laboratory, Lemont, IL}
\affil[4]{Dow Chemical Co., Midland, MI}
\affil[5]{Department of Material Science and Engineering, Cornell University, Ithaca, New York}
\affil[6]{School of Integrative Engineering, Chung-Ang University, Seoul, Republic of Korea}
\affil[7]{Department of Physics, Cornell University, Ithaca, NY}
\affil[8]{The Rowland Institute at Harvard, Cambridge, MA}
\affil[9]{Department of Electrical Engineering and Computer Science, University of Michigan, Ann Arbor, MI}
\affil[10]{National Center for Electron Microscopy, Molecular Foundry,
Lawrence Berkeley National Laboratory, Berkeley, CA}
\affil[11]{Applied Physics Program, University of Michigan, Ann Arbor, MI}

\affil[*]{e-mail: hovden@umich.edu}
\date{\today}

\abstract{Efforts to map atomic-scale chemistry at low doses with minimal noise using electron microscopes are fundamentally limited by inelastic interactions. Here, fused multi-modal electron microscopy offers high signal-to-noise ratio (SNR) recovery of material chemistry at nano- and atomic- resolution by coupling correlated information encoded within both elastic scattering (high-angle annular dark field (HAADF)) and inelastic spectroscopic signals (electron energy loss (EELS) or energy-dispersive x-ray (EDX)). By linking these simultaneously acquired signals, or modalities, the chemical distribution within nanomaterials can be imaged at significantly lower doses with existing detector hardware. In many cases, the dose requirements can be reduced by over one order of magnitude. This high SNR recovery of chemistry is tested against simulated and experimental atomic resolution data of heterogeneous nanomaterials.}

\begin{document}

\maketitle

\section*{Introduction}

Modern scanning transmission electron microscopes (STEM) can focus sub-angstrom electron beams on and between atoms to quantify structure and chemistry in real space from elastic and inelastic scattering processes. The chemical composition of specimens is revealed by spectroscopic techniques produced from inelastic interactions in the form of energy dispersive X-rays (EDX)~\cite{alfonso2010atomicEDX,kothleitner2014quantEDX} or electron energy loss (EELS)~\cite{spence1982eels, muller2008atomicEELS}. Unfortunately, high-resolution chemical imaging requires high doses (e.g., >10$^6$ e/Å$^2$) that often exceed the specimen limits---resulting in chemical maps that are noisy or missing entirely~\cite{hart2017direct,cueva2012csi}. Substantial effort and cost to improve detector hardware has brought the field closer to the measurement limits set by inelastic processes~\cite{mcmullan2014deDetectors, kotula2012chemiSTEM}. Direct interpretation of atomic structure at higher-SNR is provided by elastically scattered electrons collected in a high-angle annular dark field detector (HAADF); however, this signal under-describes the chemistry~\cite{lebeau2008quantitative}. Reaching the lowest doses at the highest SNR ultimately requires fusing both elastic and inelastic scattering modalities.


Currently, detector signals---such as  HAADF and EDX/EELS---are analyzed separately for insight into structural, chemical, or electronic properties~\cite{su2010correlativeImg}.  Correlative imaging disregards shared information between structure and chemistry and misses opportunities to recover useful information. Data fusion, popularized in satellite imaging, goes further than correlation by linking the separate signals to reconstruct new information and improve measurement accuracy~\cite{hall1997introduction,lahat2015datafusionreview,wendy2016jointXray}. Successful data fusion designs an analytical model that faithfully represents the relationship between modalities, and yields a meaningful combination without imposing any artificial connections~\cite{calhoun2016mriFusion}. 






\begin{figure*}[ht!]
    \centering
    \includegraphics[width=\linewidth]{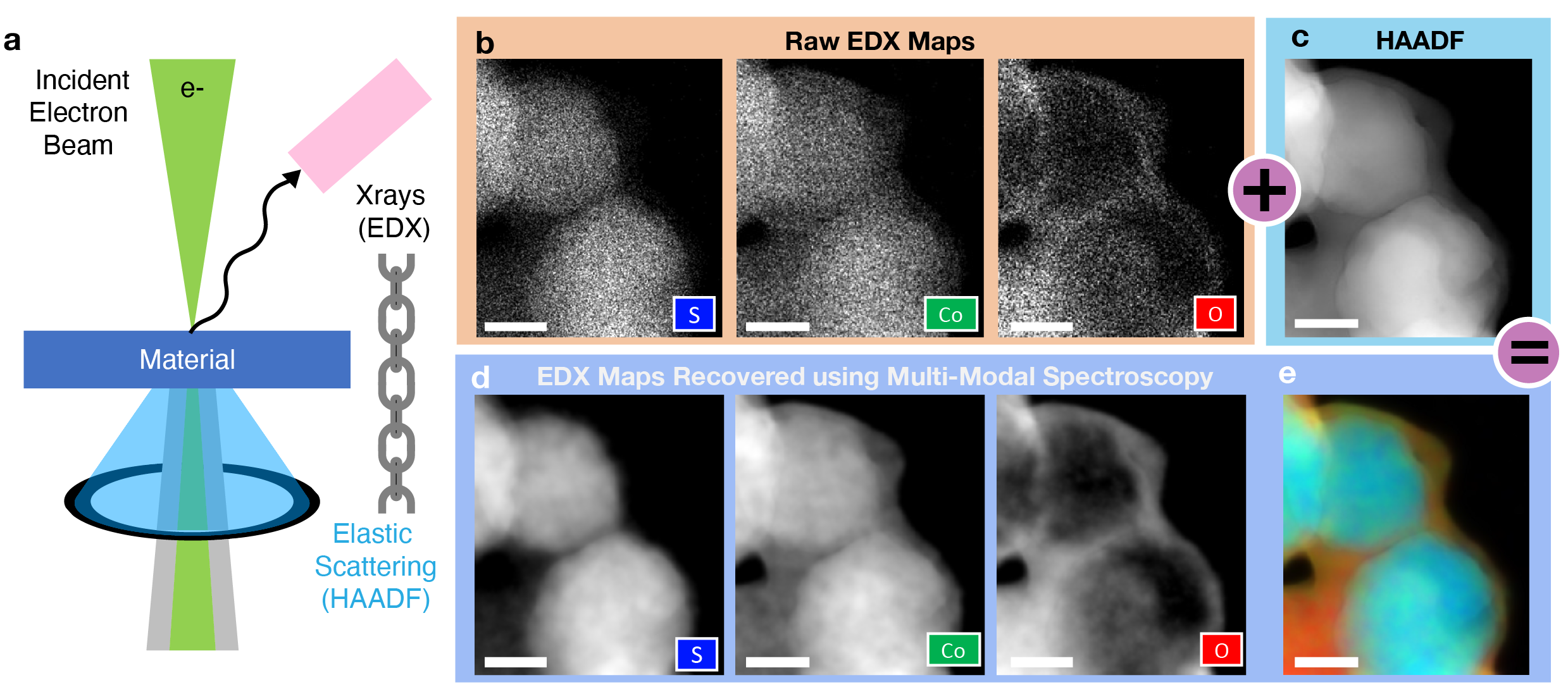}
    \caption{\textbf{Nanoscale multi-modal chemical recovery of CoS catalysts using EDX + HAADF.} \textbf{a)} Schematic highlighting the linked HAADF and EDX modalities collected in the microscope for every probe position. The algorithm links and correlates information between the two signals through an optimization process that produces chemical maps with higher SNRs. \textbf{b)} The raw EDX chemical maps for the Co, S, and O elemental distributions. \textbf{c)} The simultaneous HAADF micrograph of the CoS nanoparticle. \textbf{d)} The multi-modal reconstructions for the elemental distributions. \textbf{e)} EDX RGB overlay of the Co, S, and O maps. Scale bar, 30 nm.}
    \label{fig::edx_overview}
\end{figure*}

Here we introduce fused multi-modal electron microscopy, a technique offering high SNR recovery of nanomaterial chemistry by linking correlated information encoded within both HAADF and EDX / EELS. We recover chemical maps by reformulating the inverse problem as a nonlinear optimization which seeks solutions that accurately match the actual chemical distribution in a material. Our approach substantially improves SNRs for chemical maps, often around 300-500$\%$, and can reduce doses over one order of magnitude while remaining consistent with original measurements. We demonstrate on EDX/EELS datasets at sub-nanometer and atomic resolution. Moreover, fused multi-modal electron microscopy recovers a specimen’s relative concentration, allowing researchers to measure local stoichiometry with less-than $15\%$ error without any knowledge of the inelastic cross sections. Convergence and uncertainty estimates are identified along with simulations that provide ground-truth assessment of when and how this approach can fail.


\section*{Results}
\subsection*{Principles of Multi-Modal Electron Microscopy}

Fused multi-modal electron microscopy recovers chemical maps by solving an optimization problem seeking a solution that strongly correlates with (1) the HAADF modality containing high SNR, (2) the chemically sensitive spectroscopic modality (EELS and / or EDX), and (3) encourages sparsity in the gradient domain producing solutions with reduced spatial variation. The overall optimization function results as following:
\begin{align}
\label{eq:costFunc}
     \argminop_{\bm{x}_i \geq 0} \quad &\frac{1}{2} \Big\| \bm{b}_{H} - \sum_{i} (Z_i\bm{x}_{i})^{\gamma} \Big \|_2^2 +
     \nonumber\\
     \lambda_1 \sum_{i} & \Big(\bm{1}^T \bm{x}_i - \bm{b}_{i}^T \log(\bm{x}_i + \varepsilon) \Big) + \lambda_2 \sum_{i} \|\bm{x}_i\|_{\mathrm{TV}},
\end{align} 
where $\lambda$ are regularization parameters, $\bm{b}_H$ is the measured HAADF,
$\bm{b}_i$ and $\bm{x}_i$ are the measured and reconstructed chemical maps for element $i$,
$\varepsilon$ herein prevents log(0) issues but can also account for background,
the $\log$ is applied element-wise to its arguments,
superscript $T$ denotes vector transpose,
and
$\bm{1}$ denotes the vector of $n_x n_y$ ones,
where $n_x \times n_y$ is the image size.

The three terms in (\ref{eq:costFunc}) define our multi-modal approach to surpass traditional dose limits for chemical imaging. First, we assume a forward model where the simultaneous HAADF is a linear combination of elemental distributions ($\bm{x}_i^\gamma$ where $\gamma \in$ [1.4, 2]). The incoherent linear imaging approximation for elastic scattering scales with atomic number as $Z_i^\gamma$
where $\gamma$ is typically around 1.7~\cite{hartel1996conditions,krivanek2010atom,hovden2012efficient}.
This $\gamma$ is bounded between 2 for Rutherford scattering from bare nuclear potentials to 4/3
as described by Lenz-Wentzel expressions for electrons experiencing a screened coulombic potential~\cite{crewe1970coloumb, langmore1973eScatter}.
Second, we ensure the recovered signals maintain a high-degree of data fidelity with the initial measurements by using maximum negative log-likelihood for spectroscopic measurements dominated by low-count Poisson statistics~\cite{wendy2017jointXray,odstrcil2018iterativeLS}. In a higher count regime, this term can be substituted with a simple least-squares error. Lastly, we utilize channel-wise total variation (TV) regularization to enforce a sparse gradient magnitude, which reduces noise by promoting image smoothness while preserving sharp features~\cite{osher1992tv}. This sparsity constraint, popularized by the field of compressed sensing (CS), is a powerful yet minimal prior toward recovering structured data~\cite{donoho2006CS,candes2006CS}. When implementing, each of these three terms can and should be weighted by an appropriately selected coefficients that balances their contributions. All three terms are necessary for accurate recovery (Supplementary Figure 1).


\subsection*{High-SNR Recovery of Nanomaterial Chemistry}

Figure~\ref{fig::edx_overview} demonstrates high-SNR recovery for EDX signals of commercial cobalt sulfide (CoS) nano-catalysts for oxygen-reduction applications---a unique class with the highest activity among non-precious metals~\cite{steve2020CoSx}.
Figure \ref{fig::edx_overview}a illustrates the model that links the two modalities (EDX and HAADF) simultaneously collected in the electron microscope.
The low detection rate for characteristic X-rays is due to minimal emission
(e.g., over 50\% for $Z>32$ and below 2\% for $Z<11$) and collection yield ($<9\%$)~\cite{scholssmacher2010edxYield}.
For high-resolution EDX, the low count rate yields a sparse chemical image dominated by shot noise (Fig.~\ref{fig::edx_overview}b). However, noise in the fused multi-modal chemical map is virtually eliminated (Fig.~\ref{fig::edx_overview}d) and recovers chemical structure without a loss of resolution---including the nanoparticle core and oxide shell interface. The chemical maps produced by fused multi-modal EM quantitatively agree with the expected stoichiometry---the specimen core contains a relative concentration of 39$\pm$1.6$\%$, 42$\pm$2.5$\%$ and 13$\pm$2.4$\%$ and exterior shell composition of 26$\pm$2.8$\%$, 11$\pm$2.0$\%$, 54$\pm$1.3$\%$ for Co, S, O respectively.
The dose for this dataset was approximately $\sim$10$^5$~e Å$^{-2}$ and a 0.7 sr EDX detector was used; however, these quantitative estimates remained consistent when the dose was reduced to $\sim$10$^4$~e Å$^{-2}$.




\begin{figure*}[ht!]
    \centering
    \includegraphics[width=\linewidth]{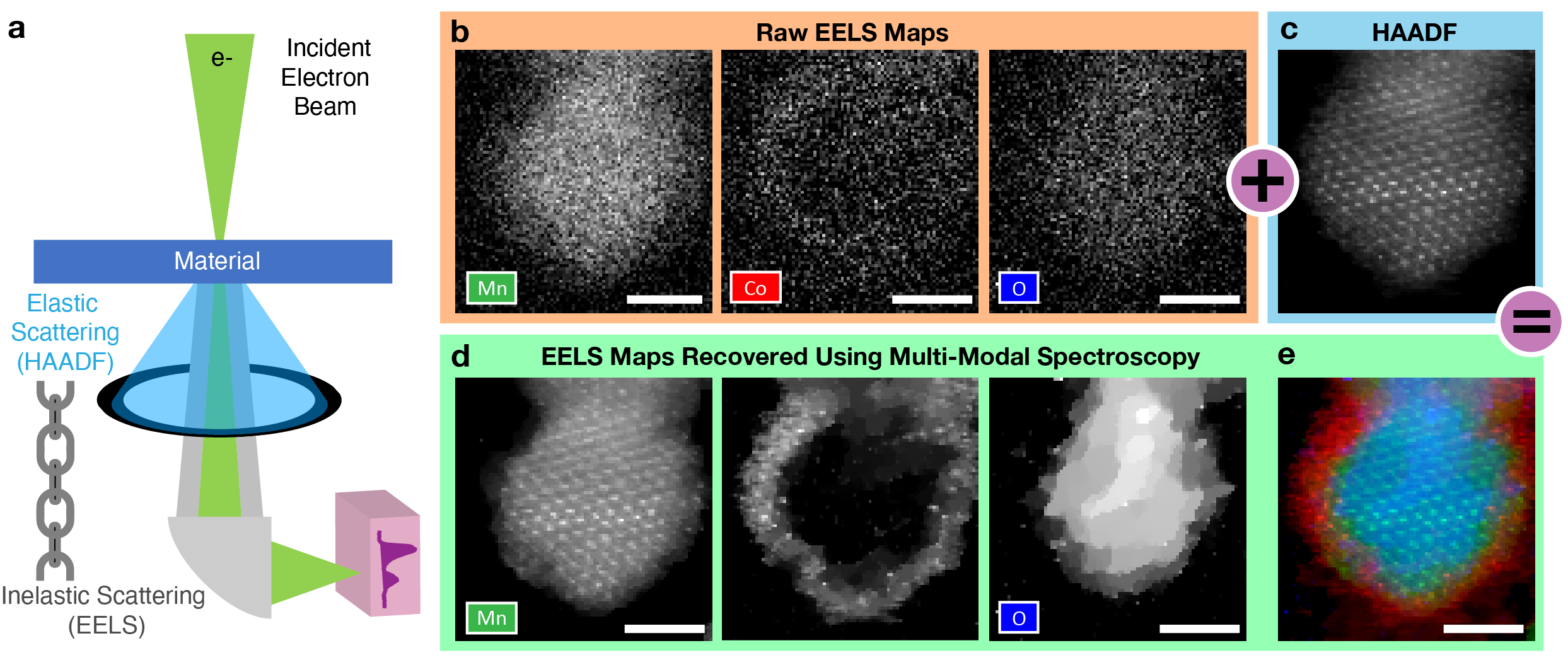}
    \caption{\textbf{Atomic-scale multi-modal chemical recovery of Co$_{3-x}$Mn$_x$O$_4$ supercapacitors using EELS + HAADF.} \textbf{a)} Schematic highlighting the linked HAADF and EELS modalities collected in the microscope at every probe position. \textbf{b)} Raw EELS maps for the elemental distributions of Co, Mn - L$_{2,3}$ and O - K edges. \textbf{c)} The simultaneous HAADF micrograph of the Co$_{3-x}$Mn$_x$O$_4$ nanoparticle. \textbf{d)} The multi-modal reconstructions for the elemental distributions. \textbf{e)} EELS RGB overlay of the Co, S, and O maps. Scale bar, 2 nm.}
    \label{fig::eels_overview}
\end{figure*}

Fused multi-modal electron microscopy accurately recovers chemical structure down to atomic length scales---demonstrated here for EELS spectroscopic signals. EELS derived chemical maps for Co$_{3-x}$Mn$_x$O$_4$ ($x=1.49$) high-performing super-capacitor nanoparticles~\cite{perera2015comno} are substantially improved by fused multi-modal electron microscopy in Figure~\ref{fig::eels_overview}. This composite Co-Mn oxide was designed to achieve a synergy between cobalt oxide's high specific capacitance and manganese oxide's long life cycle~\cite{perera2015comno,bhargava2019comnoSpinel}. While the  Co$_{3-x}$Mn$_x$O$_4$ nanoparticle appears chemically homogeneous in the HAADF projection image along the [100] direction (Fig.~\ref{fig::eels_overview}c), core-shell distinctions are hinted at in the raw EELS maps (Fig.~\ref{fig::eels_overview}b). Specifically, these nanoparticles contain a Mn-rich center with a Co shell and homogeneous distribution of O. However the raw EELS maps are excessively degraded by noise, preventing analysis beyond rough assessment of specimen morphology. The multi-modal reconstructions (Fig.~\ref{fig::eels_overview}d) confirm the crystalline Co-rich shell and map the Co/Mn interface in greater detail (Fig.~\ref{fig::eels_overview}e). In the presence of cobalt and manganese, the HAADF image lacks noticeable contrast from oxygen; the resulting oxygen map lacks detail and benefits mostly from regularization.

\begin{figure}[ht!]
    \centering
    \includegraphics[width=0.95\linewidth]{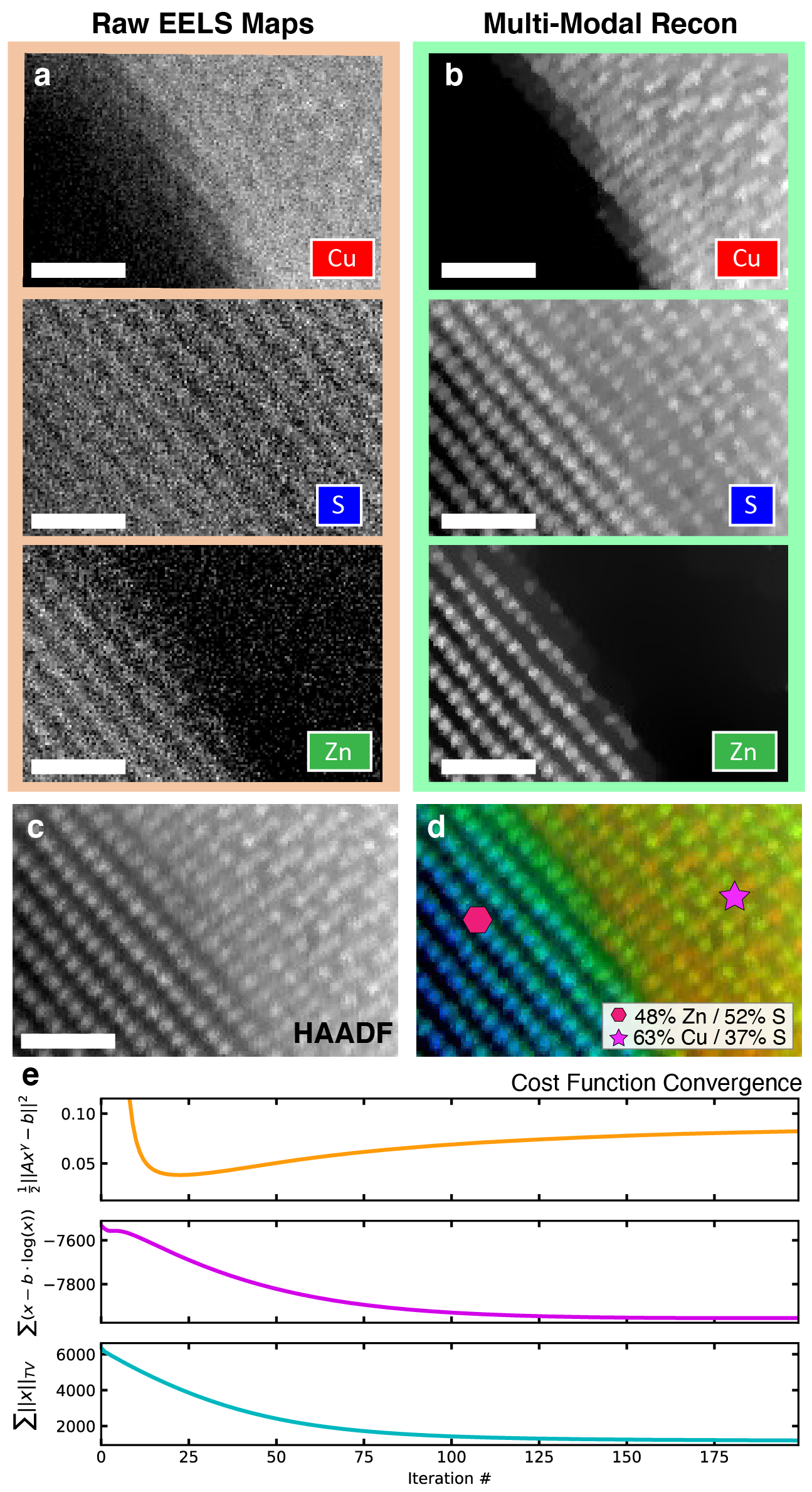}
    \caption{\textbf{Recovering chemistry in an atomically sharp ZnS-Cu$_{0.64}$S$_{0.36}$ heterointerface interface.} \textbf{a)} The raw EELS maps for the Cu, S, and Zn L$_{2,3}$ edges. \textbf{b)} The multi-modal reconstructions for the elemental compositions. \textbf{c)} The simultaneous HAADF micrograph of the ZnS-Cu$_{0.64}$S$_{0.34}$ interface. \textbf{d)} Color overlay of the Zn, S, and Zn maps. The relative concentration for the constituent elements consist of 48$\pm$5.9$\%$ for Zn, 59.9$\pm$3.2$\%$ for Cu and 38$\pm$2.6$\%$ for S in the Cu$_{0.64}$S$_{0.36}$ layer and 48.9$\pm$6$\%$ in ZnS. \textbf{e)} Convergence plots for the three individual components in the cost function. Scale bar, 1 nm.}
    \label{fig::ZnSCu}
\end{figure}

 Figure \ref{fig::ZnSCu} exhibits fused multi-modal electron microscopy at atomic resolution on copper sulphur heterostructured nanocrystals with zinc sulfide caps with potential applications in photovoltaic devices or battery electrodes~\cite{ha2014ZnSCu}. The copper sulfide properties are sensitive to the Cu-S stoichiometry and crystal structure at the interface between ZnS and Cu$_{0.64}$S$_{0.36}$. Figure \ref{fig::ZnSCu} shows high-resolution HAADF and EELS characterization of a heterostructure Cu$_{0.64}$S$_{0.36}$-ZnS interface. Fused multi-modal electron microscopy maps out the atomically sharp Cu$_{0.64}$S$_{0.36}$-ZnS interface and reveals step edges between the two layers. The labeled points on the RGB chemical overlay (Fig.~\ref{fig::ZnSCu}d) shows the chemical ratios produced by multi-modal EM for the Cu$_{0.64}$S$_{0.36}$ and ZnS regions---values which are consistent with the reported growth conditions. Figure \ref{fig::ZnSCu}e shows the algorithm convergence for each of the three terms in the optimization function (Eq.~\ref{eq:costFunc})---smooth and asymptotic decay is an indicator of reliable reconstruction. Refer to Supplementary Figure 2 for an additional demonstration at the atomic-scale on an ordered manganite system. 

\begin{figure}[ht!]
    \centering
    \includegraphics[width=\linewidth]{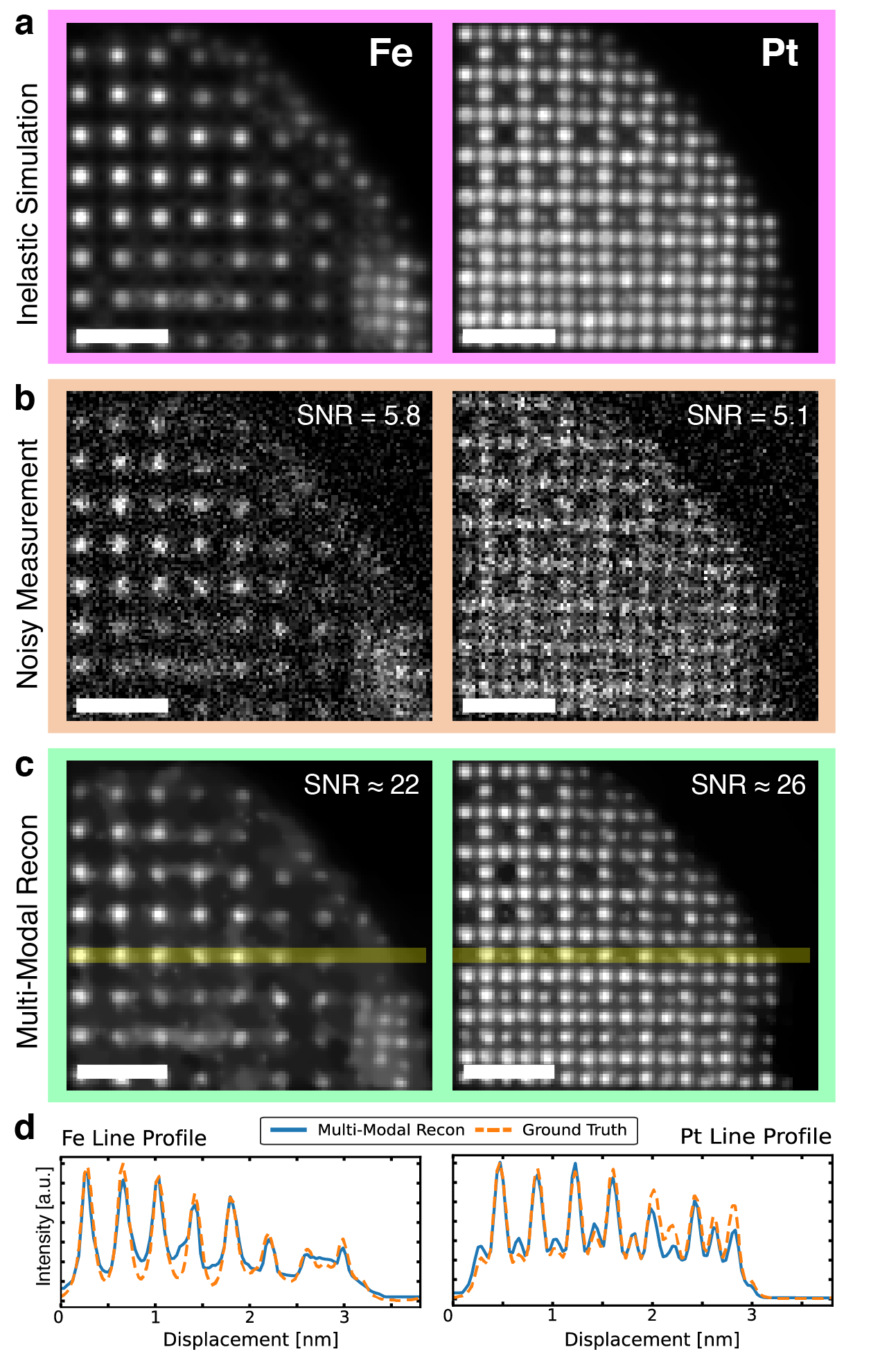}
    \caption{\textbf{Inelastic FePt nanoparticle simulation.} \textbf{a)} Ground truth EELS images generated from inelastic simulations. \textbf{b)} EELS maps degraded with Poisson shot noise. SNR shown on top right. \textbf{c)} Recovered atomic-resolution EELS maps for the Fe and Pt distributions. Estimated SNR shown on top right. \textbf{d)} Line profiles of the marked yellow bars (10 pixels in width) in (c) compares the Multi-Modal reconstruction and ground truth. Scale bar, 1 nm. }
    \label{fig::FePtSim}
\end{figure}

Fused multi-modal imaging of Fe and Pt distributions from inelastic multislice simulations (Fig.~\ref{fig::FePtSim}) provide ground truth solutions to validate recovery at atomic resolution under multiple scattering conditions of an on-axis $\sim$8 nm nanoparticle. Here, we applied Poisson noise (Fig.~\ref{fig::FePtSim}b) containing electron doses of $\sim$10$^9$~e~Å$^{-2}$, to produce chemical maps with noise levels resembling experimental atomic-resolution EELS datasets (SNR $\simeq 5$). We estimated SNR improvements by measuring peak-SNR for the noisy and recovered chemical maps \cite{hore2010psnr}. Qualitatively, the recovered chemical distributions (Fig.~\ref{fig::FePtSim}c) match the original images. Fig.~\ref{fig::FePtSim}d illustrates agreement of the line profiles as the atom column positions and relative peak intensities between the ground truth and multi-modal reconstruction are almost identical. 

Simulating EELS chemical maps is computationally demanding as every inelastic scattering event requires propagation of an additional wavefunction~\cite{dwyer2015role,allen2015modelling}---scaling faster than the cube of the number of beams, $O(N^3 \log N)$. Inelastic transition potentials of interest (in this case the L$_{2,3}$ Fe and M$_{4,5}$ Pt edges) were calculated from density function theory (see Methods). Long computation times (nearly 4,000 core-hours) result from a large number of outgoing scattering channels corresponding to the many possible excitations in a sample. For this reason, there is little precedence for inelastic image simulations. We relaxed the runtime by utilizing the PRISM STEM-EELS approximation, achieving over a ten-fold speedup (see Methods)~\cite{brown2019prismEELS}. Future work may explore the effects of smaller ADF collection angles with increased coherence lengths and crystallographic contrast~\cite{hartel1996conditions,zhang2021combining}, or thicker specimens where electron channeling becomes more concerning~\cite{anstis2003limitations,hovden2012channeling}.

\subsection*{Quantifying Chemical Concentration}

\begin{figure}[ht!]
    \centering
    \includegraphics[width=\linewidth]{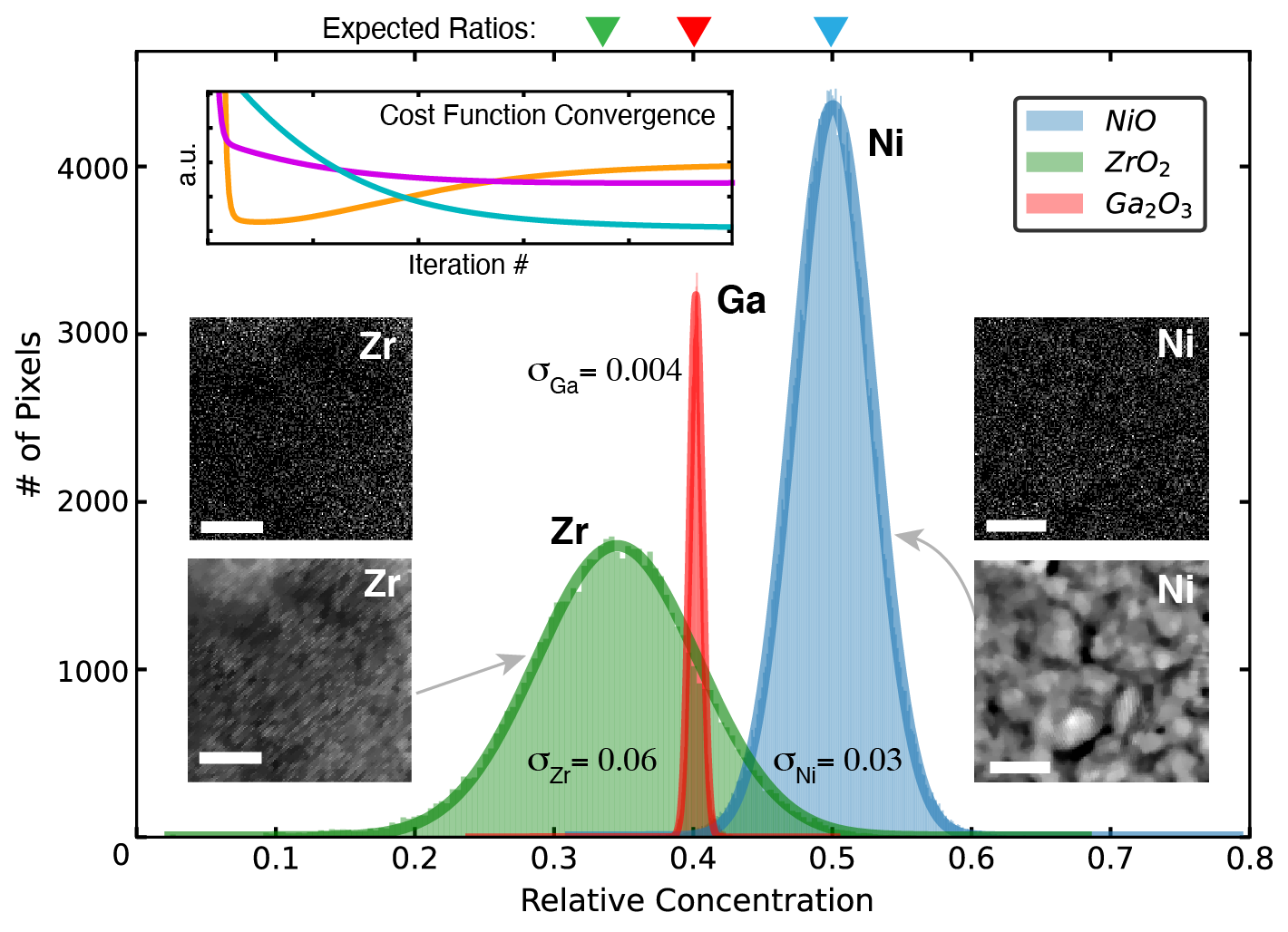} 
    \caption{\textbf{Measuring relative concentration for experimental and synthetic datasets.} Pixel intensity histograms for an experimental Zr (green), Ni (blue) and synthetic Ga (red) concentration maps. The standard deviation ($\sigma$) for each element is reported.  The raw and reconstructed EDX maps are illustrated inside of the plot. Ground truth concentrations are highlighted by the respective colored triangles above the top axis. Stable convergence for the three components in the cost function: model term (orange), data fidelity (magenta), and regularization (turquoise) are illustrated in the inset. Qualitatively the convergence is identical for all three example datasets. Zr and Ni scale bars: 5, 10 nm, respectively.}
    \label{fig::stoichiometry}
\end{figure}

Fused multi-modal electron microscopy can produce stoichiometricly meaningful chemical maps without specific knowledge of inelastic cross sections. Here, the ratio of pixel values in the reconstructed maps quantify elemental concentration. We demonstrate quantifiable chemistry on experimental metal oxide thin films with known stoichiometry: NiO~\cite{egerton1994NiO} and ZrO$_{2}$. A histogram of intensities from the recovered chemical maps are fit with Gaussian distributions to determine the average concentration. The recovered pixel values highlighted in Figure \ref{fig::stoichiometry} followed a single Gaussian distribution where the Zr and Ni concentrations are centered about 35$\pm$5.8$\%$ and 50$\pm$2.9$\%$. In both cases, the average Ni and Zr relative concentration is approximately equivalent to the expected ratio from the crystal stoichiometry: 33$\%$ and 50$\%$. The CoS nanoparticle in Fig. \ref{fig::edx_overview} follows a bi-modal distribution for the core and shell phases (Supplementary Figure 5). We found measuring stoichiometry is robust across a range of $\gamma$ values close to 1.7. In cases where $\gamma$ is far off (e.g., $\gamma = 1.0$), the quantification is systematically incorrect (Supplementary Figure 6).

We further validate stoichiometric recovery on a synthetic gallium oxide crystal (Fig~\ref{fig::stoichiometry}) where two overlapping Ga and O thin films of equal thickness have a stoichiometery of Ga$_2$O$_3$. The simulated HAADF signal is proportional to $\sum_{i} (\bm{x}_i Z_{i})^{\gamma}$ where $\bm{x}_{i}$ is the concentration for element ${i}$ and $Z_{i}$ is the atomic number. As shown by the histogram, the simulated results agree strongly with the prior knowledge and successfully recovers the relative Ga concentration. The Gaussian distribution is centered about 40$\pm$0.4$\%$ when the ground truth is 40$\%$. The inset shows convergence plots. 

We estimate a stoichiometric error of less-than 15\% for most materials based on the relative concentration's standard deviation ($\pm$7\%) added in quadrature with the variation of solutions ($\pm$6\%).
Although the algorithm shows stable convergence,
the overall quantitative conclusions are slightly sensitive to the selection of hyperparameters. 
We estimate incorrect selection of hyperparameters could result in variation of roughly $\pm$6$\%$ from the correct prediction in stoichiometery even when the algorithm converges (convergence shown in Supplementary Figures 8-9). This error is comparable to estimating chemical concentrations directly from EELS / EDX spectral maps from the ratio of scattering cross section against core-loss intensity~\cite{rez1982crossSections}. However, traditional approaches require accurate knowledge of all experimental parameters (e.g., beam energy, specimen-thickness, collection angles) and accurate calculation of the inelastic cross-section typically to provide errors roughly between 5-10$\%$~\cite{egerton1978eelsQuant}.

\subsection*{Influence of Electron Dose}

\begin{figure}[ht!]
    \centering
    \includegraphics[width=\linewidth]{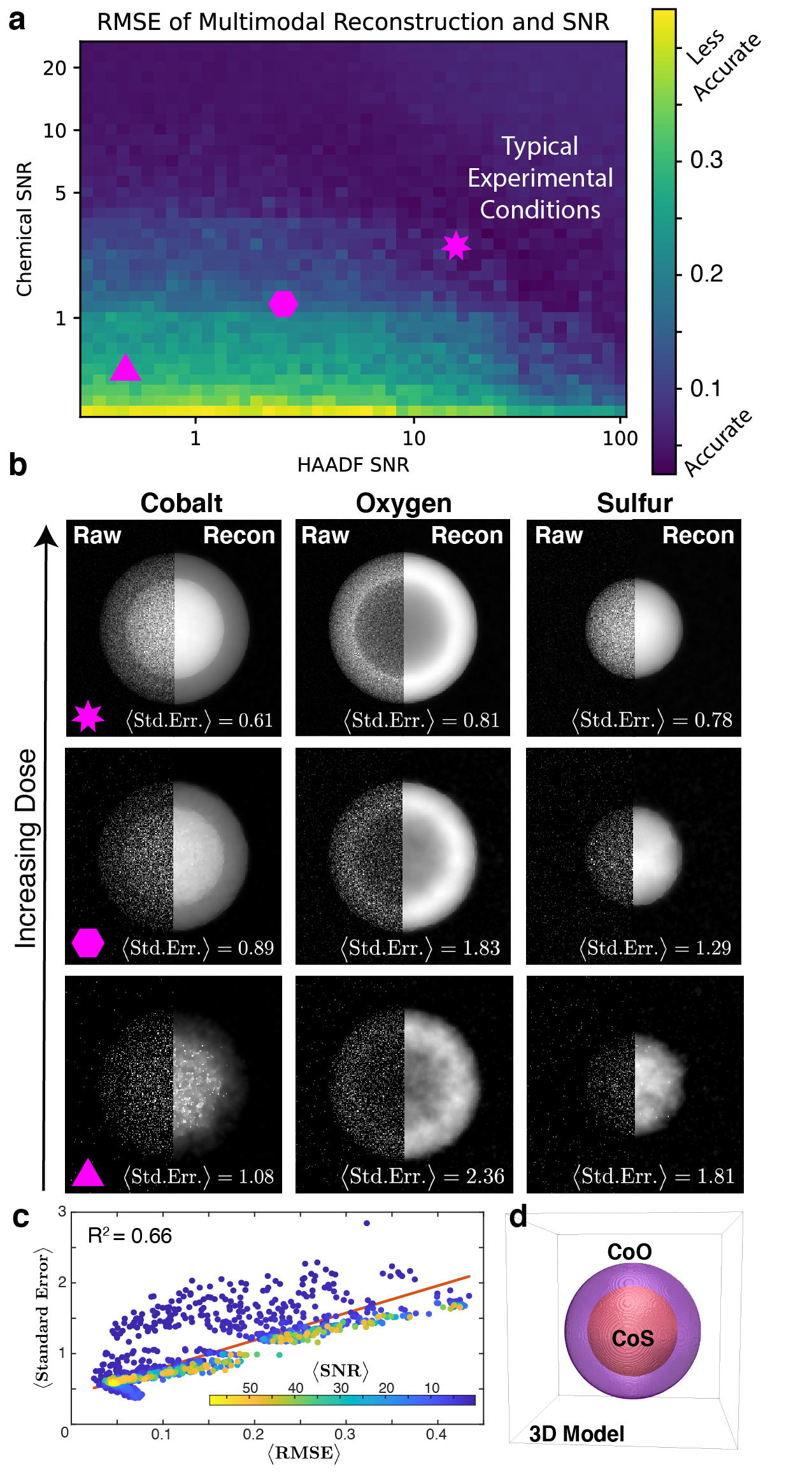}
    \caption{\textbf{Estimating dose requirements for accurate chemical recovery.} \textbf{a)} A RMSE map representing the reconstruction error as a function of multiple spectroscopic and HAADF SNR. Brighter pixels denote results containing the incorrect concentrations from the ground truth. \textbf{b)} Visualization of three points on the phase diagram corresponding to increasing ADF / chemical electron dose. \textbf{c)} A plot of average standard error vs. RMSE demonstrating the two metrics are linearly correlated. \textbf{d)} The 3D model for generating synthetic chemical and ADF projections. }
    \label{fig::3Dsimulation}
\end{figure}

To better understand the accuracy of fused multi-modal electron microscopy at low doses, we performed a quantitative study of normalized root-mean-square error (RMSE) concentrations for a simulated 3D core-shell nanoparticle (CoS core, CoO shell).
Figure~\ref{fig::3Dsimulation} shows the fused multi-modal reconstruction accuracy across a wide range of HAADF and chemical SNR. The simulated projection images were generated by simple linear incoherent imaging model of the 3D chemical compositions highlighted in Fig.~\ref{fig::3Dsimulation}d--here the probe's depth of focus is much larger than the object. Random Poisson noise corresponding to different electron dose levels was applied to vary the SNR across each pixel. 


Overall, the RMSE simulation map (Fig.~\ref{fig::3Dsimulation}a) shows the core-shell nanoparticle chemical maps are accurately recovered at low-doses (HAADF SNR $\gtrsim 4$ and chemical SNR $\gtrsim 2$); however, they become less accurate at extremely low doses. The RMSE map for multimodal reconstruction shows a predictably continuous degradation in recovery as signals diminish. The degraded and reconstructed chemical maps for various noise levels are highlighted in Figure~\ref{fig::3Dsimulation}b. The Co map closely mirrors the Z-contrast observed in HAADF (not shown) simply because it is the heaviest element present. Usually researchers will perform spectroscopic experiments in the top right corner of Fig.~\ref{fig::3Dsimulation}a
(e.g., HAADF SNR $>20$, chemical SNR $>3$), which for this simulation, provides accurate recovery.

In actual experiments, the ground truth is unknown and RMSE cannot be calculated to assess fused multi-modal electron microscopy. However we can estimate accuracy by calculating an average standard error of our recovered image from the Hessian of our model (see methods). The standard error reflects uncertainty at each pixel in a recovered chemical map by quantifying the neighborhood size for similar solutions (Supplementary Figure 10). The average standard error across all pixels in a fused multi-modal image provides a single value metric of the reconstruction accuracy (see Methods).  Figure \ref{fig::3Dsimulation}c shows that RMSE and average standard error correlate, especially at higher doses (SNR > 10).



\section*{Discussion}

While this paper highlights the advantages of multi-modal electron microscopy, the technique is not a black-box solution.
Step sizes for convergence and weights on the terms in the cost function (Eq.~\ref{eq:costFunc}) must be reasonably selected. This manuscript illustrates approaches to assess the validity of concentration measurements using confidence estimation demonstrated across several simulated and experimental material classes. Standard spectroscopic pre-processing methods become ever more critical in combination with multi-modal fusion. Improper background subtraction of EELS spectra or overlapping characteristic X-ray peaks that normally causes inaccurate stoichiometric quantification also reduces the accuracy of fused multi-modal imaging.

Fused multi-modal electron microscopy offers little advantage in recovering chemical maps for elements with insignificant contrast in the HAADF modality.
This property is limiting for analyzing specimens with low-Z elements in the presence of heavy elements (e.g., oxygen and lutetium). Future efforts could resolve this challenge by incorporating an additional complementary elastic imaging mode where light elements are visible, such as annular bright field (ABF)~\cite{findlay2010abf}. However in some instances, fused multi-modal electron microscopy may recover useful information for under-determined chemical signals. For example, in a Bi$_{0.35}$Sr$_{0.18}$Ca$_{0.47}$MnO$_3$ (BSCMO) system~\cite{savitzky2017bscmo}, only the Ca, Mn, and O EELS maps were obtained, yet multimodality remarkably improves the SNR of measured maps despite missing two elements (Supplementary Figure 2). 

Although fused multi-modal chemical mapping appears quite robust at nanometer or sub-nanometer resolution, we found atomic-resolution reconstructions can be challenged by spurious atom artifacts which require attention. However, this is easily remedied by down-sampling to frequencies below the first Bragg peaks and analysing a lower resolution chemical map. Alternatively, recovery with minimal spurious atom artifacts is achieved when lower resolution reconstructions are used as an initial guess (Supplementary Figure 11).

In summary, we present a model-driven data fusion algorithm that substantially improves the quality of electron microscopy spectroscopic maps at nanometer to atomic resolutions by using both elastic and inelastic signals.
From these signals, or modalities, each atom's chemical identity and coordination provides essential information about the performance of nanomaterials across a wide range of applications from clean energy, batteries, and opto-electronics, among many others. In both synthetic and experimental datasets, multi-modal electron microscopy shows quantitatively accurate chemical maps with values that reflect stoichiometry. This approach not only improves SNR but opens a pathway for low-dose chemical imaging of radiation sensitive materials. Although demonstrated herein for common STEM detectors (HAADF, EDX, and EELS), this approach can be extended to many other modalities---including pixel array detectors, annular bright field, ptychography, low-loss EELS, etc. One can imagine a future where all scattered and emitted signals in an electron microscope are collected and fused for maximally efficiently atomic characterization of matter.

\input{appendix.tex}

\section*{References}

{\printbibliography[heading=none]}

\section*{Acknowledgements}

R.H. and J.S. acknowledge support from the Army Research Office, Computing Sciences (W911NF-17-S-0002) and Dow Chemical Company.

\section*{Author Contributions}
J.S., R.H., Z.W.D. and Y.J. conceived the idea.
J.S. and R.H. implemented the multi-modal reconstruction algorithms and performed the analysis.
J.F. and Z.W.D. assisted with the algorithm formulation.
C.O. designed and ran the inelastic multi-slice simulations.
S.R. and A.J.F. conducted the EDX experiments.
R.H. and I.E.B. conducted the EELS experiments.
R.R., D. H., S.D.P., synthesized the Co$_{3-x}$Mn$_x$O and ZnSCu nanoparticles.
J.S. and R.H. wrote the manuscript.
All authors reviewed and commented on the manuscript. 

\section*{Ethics Declarations}
\subsection*{Competing Interests}
The authors declare no competing interests. 

\end{document}

%% file: preamble.tex
\usepackage{amsmath}
\usepackage{amssymb}
\usepackage{gensymb}
\usepackage{textcomp}
\usepackage{xcolor}
\usepackage{soul}
\usepackage{bm}
\usepackage[normalem]{ulem}
\bibliography{refs}

\newcommand{\comment}[1]{}
\long\def\red#1{\bgroup\color{red}#1\egroup}
\mathchardef\mhyphen="2D

\DeclareMathOperator*{\argminop}{arg\,min} 

%% file: appendix.tex
\section*{Methods}

\subsection*{Electron Microscopy}
Simultaneously acquired EELS and HAADF datasets were collected on a 5-th order aberration-correction Nion UltraSTEM microscope operated at 100 keV with a probe semi-angle of roughly 30 mrad and collection semi-angle of 80-240 mrad and 0-60 mrad for HAADF and EELS, respectively. Both specimens were imaged at 30 pA, for a dwell time of 10 ms (Fig. \ref{fig::ZnSCu}) and 15 ms (Fig. \ref{fig::eels_overview}) receiving a total dose of 3.25 $\times 10^4$ and 7.39 $\times 10^4$ e/Å$^2$. The EELS signals were obtained by integration over the core loss edges, all of which were done after background subtraction. The background EELS spectra were modeled using a linear combination of power laws implemented using the open-source Cornell Spectrum Imager software \cite{cueva2012csi}. 

Simultaneously acquired EDX and HAADF datasets were collected on a Thermo Fisher Scientific Titan Themis G2 at 200 keV with a probe semi-angle of roughly 25 mrad, HAADF collection semi-angle of 73-200 mrad, and 0.7 sr EDX solid angle. The CoS specimen was imaged at 100 pA and 40 $\mu$s dwell time for 50 frames receiving a total dose of approximately $2 \times 10^{5}$ e/Å$^2$. The initial chemical distributions were generated from EDX maps using commercial Velox softwarethat produced inital net count estimates (however atomic percent estimates are also suitable).

\subsection*{Fused Multi-Modal Recovery}
 

Here, fused multi-modal electron microscopy is framed as an inverse problem expressed in the following form:
$\hat{\bm{x}} = \argminop_{\bm{x} \geq 0} \Psi_1(\bm{x}) + \lambda_1 \Psi_2(\bm{x}) + \lambda_2 \mathrm{TV}(\bm{x})$
where $\hat{\bm{x}}$ is the final reconstruction, and the three terms are described in the main manuscript (Eq.~\ref{eq:costFunc}). When implementing an algorithm to solve this problem, we concatenate the multi-element spectral variables ($\bm{x}_i, \bm{b}_i)$ as a single vector:
$\bm{x},~\bm{b}~\in~\mathbb{R}^{n_x n_y n_{i}}$
where $n_{i}$ denotes the total number of reconstructed elements. 
 
 

The optimization problem is solved by a combination of gradient descent with total variation regularization. We solve this cost function by descending along the negative gradient directions for the first two terms and subsequently evaluate the isotropic TV proximal operator to denoise the chemical maps~\cite{beck2009tv}. The gradients of the first two terms are:
\begin{align}
\nabla_{\bm{x}} \Psi_1(\bm{x}) &= - \gamma \text{diag} \big(\bm{x}^{\gamma-1}\big) \bm{A}^{T} \big(\bm{b}_{H} - \bm{A} \bm{x}^{\gamma} \big) \\
\nabla_{\bm{x}} \Psi_2(\bm{x}) &= \textbf{1} - \bm{b} \oslash (\bm{x} + \varepsilon),
\end{align}
where $\oslash$ denotes point-wise division.
Here, the first term in the cost function,
relating the elastic and inelastic modalities,
has been equivalently re-written as
$\Psi_1 = \frac{1}{2} \big\| \bm{b}_{H} - \bm{A} \bm{x}^{\gamma} \big \|_2^2$,
where $\bm{A}~\in~\mathbb{R}^{n_x n_y \times n_x n_y n_i}$ expresses the summation of all elements as matrix-vector multiplication. Evaluation for the TV proximal operator is in itself another iterative algorithm. In addition, we impose a non-negativity constraint since negative concentrations are unrealistic. We initialize the first iterate with the measured data ($\bm{x}^0_i = \bm{b}_i$), an ideal starting point as it is a local minima for $\Psi_2$.

The inverse of the Lipschitz constant (1/L) is an upper bound of the step-size that can theoretically guarantee convergence. From Lipschitz continuity, we estimated the step size for the model term's gradient ($\nabla \Psi_1$) as: 1/$L_{\nabla\Psi_1} \leq 1/\big(\|\bm{A}\|_1 \|\bm{A}\|_{\infty}\big) = 1/n_i$. The gradient of the Poisson negative log-likelihood ($\Psi_2$) is not Lipschitz continuous, so its descent parameter cannot be pre-computed \cite{dupe2009poissonLips}. We heuristically determined the regularization parameters starting with values with a similar order of magnitude to $1/L_{\nabla \Psi_1}$, then iteratively reduce until the cost function exhibits stable convergence. The regularization parameters were manually selected, however future work may allow automated optimization by the L-curve method or cross-validation \cite{regParamSelect}.




\subsection*{Estimating Standard Error of Recovered Chemical Maps}

Using estimation theory,
we can approximate the uncertainty in a recovered chemical image for unbiased estimators with the model's (Eq.~\ref{eq:costFunc}) Hessian expressed as: $\bm{H}(\bm{x}) = \nabla_{\bm{x}}^2 \Psi_1(\bm{x}) + \nabla_{\bm{x}}^2 \Psi_2(\bm{x}) $, where \begin{align}
\nabla_{\bm{x}}^2 \Psi_1(\bm{x}) &= \text{diag} \Big( \gamma(\gamma - 1) \text{diag} \big(\bm{x}^{\gamma-2}\big) \bm{A}^{T} \big(\bm{b}_{H} - \bm{A} \bm{x}^{\gamma} \big) \Big) \\ & \nonumber \quad + \gamma^2 \text{diag} \big( \bm{x}^{\gamma-2} \big) \bm{A}^T \bm{A} \text{diag}(\bm{x}^{\gamma-1}) \\
\nabla_{\bm{x}}^2 \Psi_2(\bm{x}) &= \text{diag} \big( \bm{b} \oslash (\bm{x} + \varepsilon)^2 \big)
\end{align}

Calculation of standard error follows the Cramer-Rao inequality, which provides a lower bound given by:
$\big(\mathrm{\textbf{Var}}(\hat{x}_j) \geq \big[ \bm{H}^{-1}(\hat{\bm{x}}) \big]_{jj} \big)$ \cite{wei2020crlbPtycho}, where $\mathrm{\textbf{Var}}(\hat{\bm{x}})$ are variance maps for the recovered chemical distributions ($\hat{\bm{x}}$) and subscript $jj$ denotes indices along the diagonal elements. We determined this lower bound from an empirical derivation of the Fisher Information Matrix. From the variance we thus extract standard error maps: $\mathrm{\textbf{Standard~Error}} ~= \sqrt{\mathrm{\textbf{Var}}(\hat{\bm{x}})}$ as demonstrated in Supplemental Figure 10. The average standard error denotes the mean value of all pixels in $\mathrm{\textbf{Standard~Error}}$. Note, the TV regularizer reduces noise and may introduce bias due to smoothing, so the standard error measurements could potentially be lower; our Fisher information derivation provides an upper bound on uncertainty.





\subsection*{Inelastic Scattering Simulations for Atomic Imaging}
The inelastic scattering simulations for the FePt nanoparticle structure (Fig.~\ref{fig::FePtSim}) were performed using the abTEM simulation code \cite{madsen2021abTEM}, using the algorithm described in \cite{brown2019prismEELS}. In this algorithm the initial STEM probe is propagated and transmitted to some depth into the specimen using the scattering matrix method described in the PRISM algorithm \cite{ophus2017PRISM}. Next, the inelastic transition potentials of interest (in this case the L$_{2,3}$ Fe and M$_{4,5}$ Pt edges) were calculated and applied using the methods given in \cite{saldin1987theoryEELS, dwyer2008elasticScatter}, using the GPAW density functional theory code \cite{enkovaara2010gpaw}. Finally, a second scattering matrix is used to propagate the inelastically scattered electrons through the sample and to the plane of the EELS entrance aperture. The elastic signal channels were calculated with the conventional PRISM method using the same parameters.

The atomic structure used in the simulations was a portion of the FePt nanoparticle structure determined from atomic electron tomography \cite{yang2017feptExp}. After cropping out 1/4 of nanoparticle coordinates, the boundaries were padded by 5 Å total vacuum. The STEM probe's convergence semiangle was set to 20 mrad and the voltage to 200 kV. The multislice steps used a slice thicknesses of 2 Å, the wavefunction sampling size was 0.15 Å, and the projected potentials were computed using the infinite Kirkland parameterization \cite{kirkland2020compuTEM}. The EELS detector had a semiangle of 30 mrad, and the STEM probe positions were Nyquist sampled at a step size of 0.31 Å. After completion, we convolved the simulated images with a 0.2 Å Gaussian to account for source size. These simulation parameters required approximately 4 days of calculation time using the CPU mode of abTEM on a workstation with a 40 core Xeon processor clocked at 2.0 GHz.

\section*{Data Availability}
The datasets and codes that support the finding of this study are available from the corresponding author upon reasonable request. 